\documentclass[english,floatfix,twocolumn,prl,aps]{revtex4}
\usepackage[T1]{fontenc}
\usepackage[latin1]{inputenc}
\usepackage{graphicx}
\usepackage{amssymb}
\usepackage{amsmath}
\usepackage{bm}
\usepackage{babel}
\usepackage{color}
\makeatother

\begin{document}

\title{Uniaxial Strain in Graphene by Raman Spectroscopy: G peak splitting, Gruneisen Parameters and Sample Orientation}

\author{T. M. G. Mohiuddin$^1$, A. Lombardo$^2$, R. R. Nair$^1$, A. Bonetti$^2$, G. Savini$^2$, R. Jalil$^1$\\ N. Bonini$^3$, D.M. Basko $^4$, C. Galiotis$^5$, N. Marzari$^3$, K.
S. Novoselov$^1$,\\ A. K. Geim$^1$, A. C. Ferrari$^2$}\email{acf26@eng.cam.ac.uk}

\affiliation{$^1$Department of Physics and Astronomy, Manchester University, Manchester UK\\
$^2$Engineering Department, Cambridge University, Cambridge, UK\\
$^3$Department of Materials Science and Engineering, Massachusetts Institute of Technology, Cambridge, MA, USA\\
$^4$Laboratoire de Physique et Mod\'elisation des Mileux Condens\'es,
Universit\'e Joseph Fourier and CNRS, Grenoble, France\\
$^5$ FORTH/ ICEHT and Materials Science Dept., University of Patras, Patras, Greece}

\begin{abstract}  Graphene is the two-dimensional building block for carbon allotropes of every other dimensionality. Since its
experimental discovery, graphene continues to attract enormous interest, in particular as a new
kind of matter, in which electron transport is governed by a Dirac-like wave equation, and as a
model system for studying electronic and phonon properties of other, more complex, graphitic
materials\cite{Nov306(2004),CastroNetoRev,Nov438(2005),Zhang438(2005)}. Here, we uncover the
constitutive relation of graphene and probe new physics of its optical phonons, by studying its
Raman spectrum as a function of uniaxial strain. We find that the doubly degenerate E$_{2g}$
optical mode splits in two components, one polarized along the strain and the other perpendicular
to it. This leads to the splitting of the G peak into two bands, which we call G$^+$ and G$^-$, by
analogy with the effect of curvature on the nanotube G peak\cite{piscaPRB,kresse,jorio}. Both peaks
red shift with increasing strain, and their splitting increases, in excellent agreement with
first-principles calculations. Their relative intensities are found to depend on light
polarization, which provides a useful tool to probe the graphene crystallographic orientation with
respect to the strain. The singly degenerate 2D and 2D' bands also red shift, but do not split for
small strains. We study the Gruneisen parameters for the phonons responsible for the G, D and D'
peaks. These can be used to measure the amount of uniaxial or biaxial strain, providing a
fundamental tool for nanoelectronics, where strain monitoring is of paramount
importance\cite{straingen1,straingen2}\end{abstract} \maketitle

\begin{figure}
\centerline{\includegraphics[width=89mm]{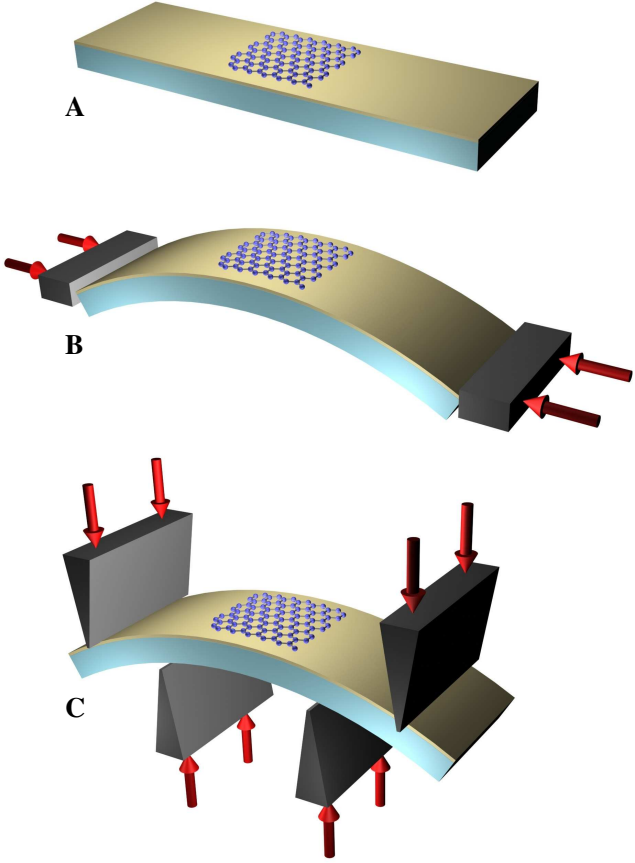}}
\caption{(Color Online) (A) Scheme (not to scale) of the substrate coated with SU8. A graphene monolayer is placed in the middle; (B,C) Scheme
(not to scale) of (B) two point, and (C) four point bending set up. Note that a typical sample is 10$^3$-10$^4$ smaller
than the substrate length, see Methods.}
\label{Figure1}
\end{figure}
\begin{figure}
\centerline{\includegraphics[width=90mm]{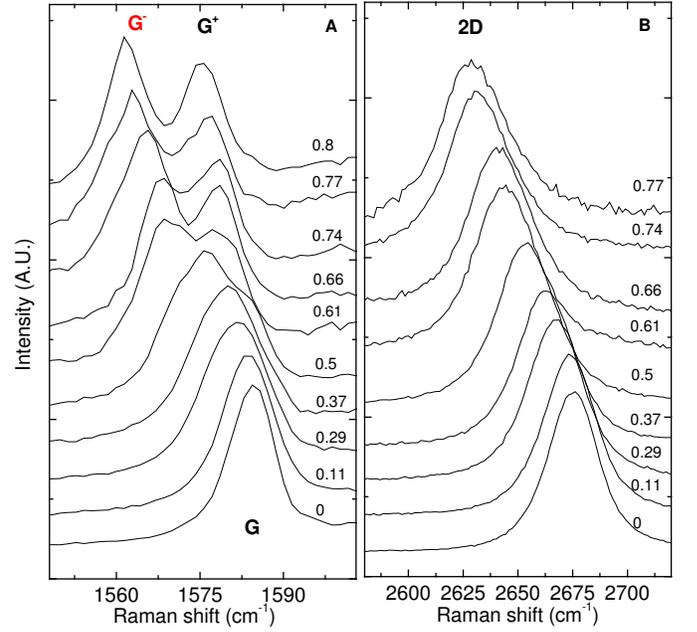}}
\caption{(Color Online) (a) G (b) 2D peaks as a function of uniaxial strain. The spectra are measured with incident light polarized along the strain direction,
collecting the scattered light with no analyzer, see Methods. Note that the doubly degenerate G peak splits in two subbands, G$^+$ and G$^-$, while this
does not happen for the 2D peak. The strains, ranging from 0 to $\sim$0.8\%, are indicated on the right side of the spectra}
\label{Figure2}
\end{figure}
\begin{figure}
\centerline{\includegraphics[width=90mm]{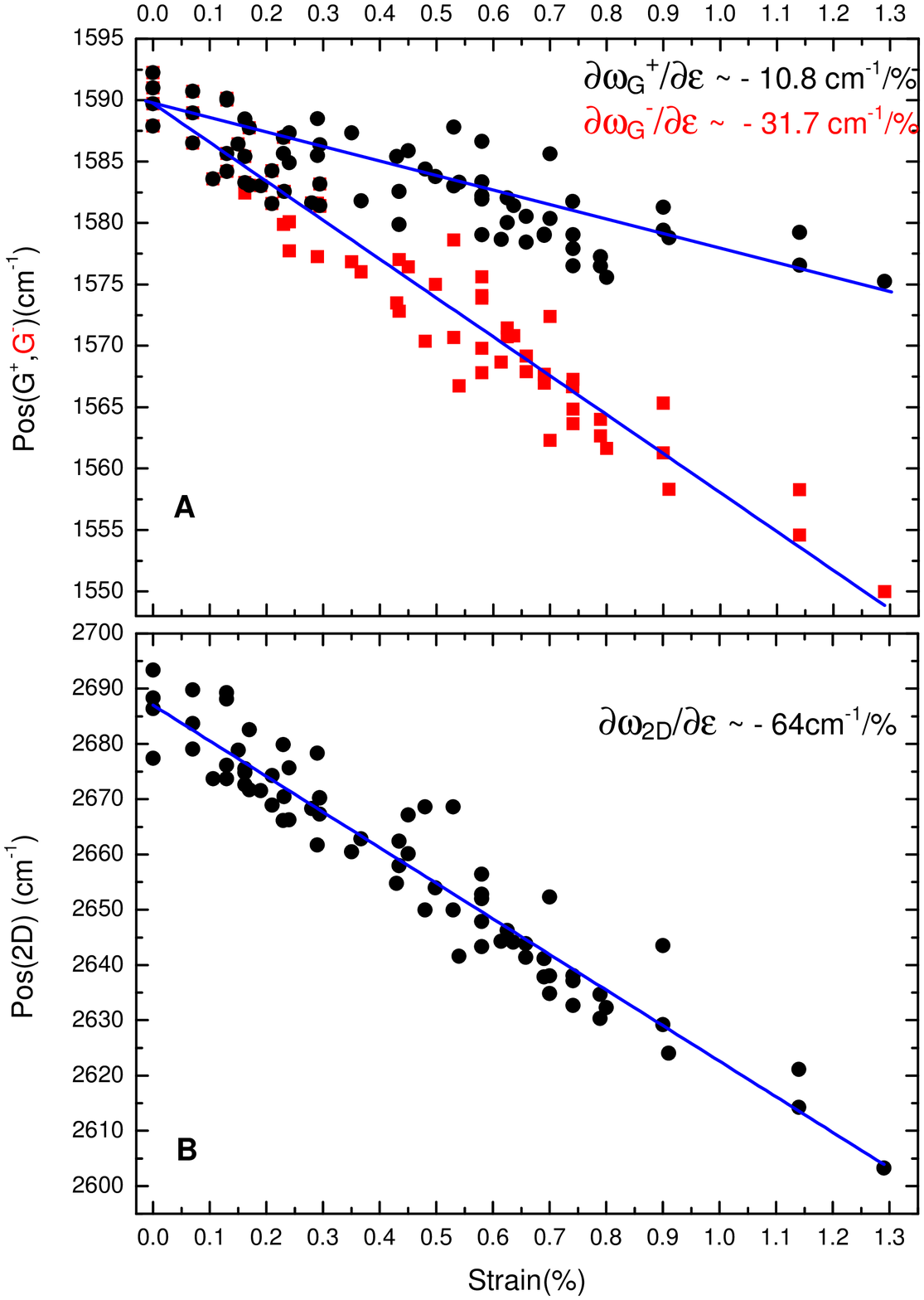}}
\caption{(Color Online) Positions (Pos) of the (a) G$^+$ and G$^-$, and (b) 2D peaks, as a function of
applied uniaxial strain. The blue lines are linear fits
to the data. The slopes of the fitting lines are also indicated}
\label{Figure3}
\end{figure}

Strain arises when a crystal is compressed or stretched out of its equilibrium shape, with the
stiffness tensor providing the constitutive relation between applied stress and final strain state.
Atomic relaxations often accompany the process, also resulting in an effective renormalization of
the constitutive relations. The presence of strain can significantly affect device performance.
Sometimes, strain is intentionally applied to improve mobility, as in the strained silicon
technology, which is used in modern microelectronics. Thus, the precise determination and
monitoring of stress and strain is a key requirement\cite{straingen1,straingen2}. Strain modifies
the crystal phonons, with tensile strain usually resulting in mode softening, and the opposite for
compressive. The rate of these changes is summarized in the Gruneisen parameters, which also
determine the thermomechanical properties\cite{grimvall}. Thus, monitoring phonons is often the
clearest and simplest way to detect strain and, if the Gruneisen parameters are known, to quantify
it.

Raman spectroscopy has emerged as the main technique to probe graphene's phonons\cite{ACFRaman}. It
can identify the number of layers in a sample\cite{ACFRaman}, determine the amount of doping and
presence of disorder\cite{Pisana,ACFRamanSSC,DasCM,CasiraghiAPL,Leandro}, study graphene's
edges\cite{cancado08,cancado04,Casiraghiedge,edgeyou,stampfer} and quantify anharmonic processes
and thermal conductivity\cite{bonini07,balandin}. Raman studies of graphene also revealed novel
physical phenomena, such as Kohn anomalies\cite{PiscanecPRL}, and the breakdown of the
Born-Oppenheimer approximation\cite{Pisana,DasCM,dasbi,Yan}. In all these cases, experimental
observations have successfully partnered with first-principles calculations, the latter providing
additional microscopic insights and understanding, while being validated by the comparison with
measurements. The Gruneisen parameters for the vibrational modes of graphite under biaxial strain
were calculated by first-principles, yielding excellent agreement with the thermomechanical
properties of graphite\cite{mounet}. Recently, changes to the Raman spectra were reported due to
the presence of stress in graphene\cite{ni,robinson,ferralis, ni2, yu, seyller}, but the inferred
strains disagreed by a factor 5 or more for similar Raman shifts\cite{ni,ferralis, ni2,yu}.
Furthermore, no significant difference was seen between the cases of uniaxial and biaxial
strain\cite{ni, ni2,yu}, in contrast with theory, and the opening of a band gap at the K point was
suggested\cite{ni}, again in contrast with theory for small strains. It is thus necessary to
conduct an accurate study in order to uncover the physics of strain for the graphene phonons.

In this work, we carefully apply uniaxial strain up to $\sim$1.3\% to a graphene monolayer, in
typical steps of 0.05\% (minimum step 0.01\%; maximum 0.25\%) using two and four point bending
setups as described in Methods (see Fig. 1), and compare this with first-principles calculations.
The Raman spectra measured at each step are fully reproducible over multiple loading and unloading
cycles, with no hysteresis. This allows us to clarify the picture for Raman spectra in strained
graphene.

Figure 2 plots some representative spectra as a function of strain. The origin of the main Raman
peaks is explained in Methods. The strain is parallel to the longest side of the substrate (Fig 1),
and is given by the ratio of substrate thickness to twice the radius of curvature. The spectra are
fitted with lorentzians, and Fig. 3 plots the resulting trends for the G and 2D peaks. Note that
Figs.3a,b are a combination of over 80 measurements on two samples, strained in two different
experimental set-ups, and include a loading, unloading and final loading cycle. Within the
spectrometer resolution we find no difference on pre-history and, for a single sample and cycle,
the strain dependence is smooth. Linear fits using all the data yield $\partial
\omega_{G^+}/\partial \varepsilon \sim$-10.8 cm$^{-1}$/ \%; $\partial \omega_{G^-}/
\partial \varepsilon \sim$-31.7 cm$^{-1}$/ \%; $\partial \omega_{2D} / \partial \varepsilon\sim$-64
cm$^{-1}$/ \% and $\partial \omega_{2D'}/ \partial \varepsilon \sim$-35 cm$^{-1}$/ \%. Where we
call G$^+$ and G$^-$ the higher and lower G sub-bands, by analogy with
nanotubes\cite{jorio,piscaPRB}.

The observed behavior can be explained by considering the effect of uniaxial strain on the optical
modes responsible for the G, and D and D' peaks, respectively.

The Gr\"uneisen parameter for the doubly-degenerate, in-plane, Raman active $E_{2g}$ phonon,
$\gamma_{E_{2g}}$, is\cite{grimvall}:
\begin{equation}
\gamma_{E_{2g}}=-\frac{1}{\omega^0_{E_{2g}}}\frac{\partial\omega^h_{E_{2g}}}{\partial\varepsilon_{h}}
\end{equation}
where $\varepsilon_{h}=\varepsilon_{ll}+\varepsilon_{tt}$ is the hydrostatic component of the
applied uniaxial strain, $l$ is the longitudinal direction, parallel to the strain, and $t$ is the
direction transverse to it; $\omega^0_{E_{2g}}$ is the G peak position at zero strain. The shear
deformation potential $\beta_{E_{2g}}$ is defined as\cite{Reich,Thomsen}:
\begin{equation}
\beta_{E_{2g}}=\frac{1}{\omega^0_{E_{2g}}}\frac{\partial\omega^s_{E_{2g}}}{\partial\varepsilon_{s}}
\end{equation}
where $\varepsilon_{s}=\varepsilon_{ll}-\varepsilon_{tt}$ is the shear component of the strain.

Under uniaxial strain, the solution of the secular equation for the $E_{2g}$ mode is\cite{Cardona,
Sakata, Reich, Thomsen}:
\begin{eqnarray}
\Delta\omega^{\pm}_{E_{2g}} &=& \Delta\omega^h_{E_{2g}}\pm\frac{1}{2}\Delta\omega^s_{E_{2g}} \nonumber \\
&=& -\omega^0_{E_{2g}}\gamma_{E_{2g}}\left(\varepsilon_{ll}+\varepsilon_{tt}\right)\pm\frac{1}{2}
\beta_{E_{2g}}\omega^0_{E_{2g}}\left(\varepsilon_{ll}-\varepsilon_{tt}\right)\nonumber \\
\end{eqnarray}
where $\Delta\omega^h_{E_{2g}}$ is the shift resulting from the hydrostatic component of the
strain, and $\Delta\omega^s_{E_{2g}}$ is the mode splitting due to the shear component of the
strain. $\Delta \omega_{G^+} = \Delta\omega^+_{E_{2g}}$ and  $\Delta \omega_{G^-}=
\Delta\omega^-_{E_{2g}}$ are the shifts of the G$^+$ and G$^-$ peaks relative to zero strain.

It is important to note that the resulting phonon eigenvectors are orthogonal to each
other\cite{Cardona, Sakata, Reich, Thomsen}, with the E$^+{_{2g}}$ perpendicular to the applied
strain (and thus experiencing smaller softening), and the E$^-{_{2g}}$ parallel to it. This is
analogous to the effect of curvature on the G peak of carbon nanotubes. The G peak splitting in
nanotubes is the combined result of electron confinement and curvature\cite{piscaPRB}. Pure
curvature splits the graphene E$_{2g}$ mode in a component parallel to the tube axis and one
perpendicular. When the sp$^2$ bonds of graphene are deformed by rolling it in a tube, they
lengthen and soften in the direction perpendicular to the axis, in order for the $\pi_z$ electrons
to be perpendicular to it. This is proportional to curvature, so it is minimum parallel to the
axis, and maximum along the circumference, increasing with decreasing
diameter\cite{piscaPRB,kresse}. Thus, by curvature only, nanotubes will have a TO G$^-$ peak and a
LO G$^+$, with the former softer than the latter, and more sensitive to diameter changes. This
simple picture is reasonable for semiconducting nanotubes\cite{piscaPRB}, while in metallic, a
further significant softening of the LO mode takes place due to the enhanced Kohn anomaly resulting
from electron confinement\cite{piscaPRB}. However, this further effect must be absent in "unrolled"
tubes, i.e. graphene. Indeed, the Full Width at Half Maximum (FWHM) of the G$^+$ and G$^-$ peaks in
graphene is roughly constant as a function of strain at $\sim12 cm^{-1}$, whereas FWHM(G$^-$) in
metallic nanotubes becomes much larger, due to the increased electron-phonon coupling
contribution\cite{piscaPRB}.

By fitting the trends in Fig. 3 to Eqs. 1,2, 3 we can experimentally determine the Gruneisen
parameters for graphene. Under uniaxial strain\cite{grimvall} $\varepsilon_{ll}=\varepsilon$ and
$\varepsilon_{tt}=-\nu \varepsilon$. Where $\nu$ is the Poisson's ratio. If one could strain
free-hanging graphene samples, the Poisson's ratio for graphene itself should be used. This can be
taken as the in-plane Poisson's ratio of graphite $\sim$0.13~\cite{bosak}. However, the lack of
loading-unloading hysteresis for our results implies good adhesion between graphene and our
substrates for the whole range of applied strains. SU8 is a transversely isotropic material with a
0.33 in plane Poisson's ratio\cite{su8}. PET and perspex have also Poisson's ratios between
0.3-0.35. We thus use $\nu$=0.33. This corresponds to the case of ideal contact between graphene
and substrate. Eq. 3 is now rewritten as:
\begin{eqnarray}
\Delta\omega^{\pm}_{E_{2g}} &=& -\omega^0_{E_{2g}}\gamma_{E_{2g}}\left(1-\nu\right)\varepsilon\pm\frac{1}{2}
\beta_{E_{2g}}\omega^0_{E_{2g}}\left(1+\nu\right)\varepsilon \nonumber \\
\end{eqnarray}
yielding:
\begin{eqnarray}
\gamma_{E_{2g}} &=&  -\frac{\Delta \omega_{G^+} + \Delta \omega_{G^-}}
{2 \omega_{G_0}\left(1-\nu\right)\varepsilon }
\end{eqnarray}
\begin{eqnarray}
\beta_{E_{2g}} &=&  \frac{\Delta \omega_{G^+} - \Delta \omega_{G^-}}
{\omega_{G_0}\left(1+\nu\right)\varepsilon }
\end{eqnarray}
From the data in Fig.3a we get $\gamma_{E_{2g}}$=1.99; $\beta_{E_{2g}}$=0.99. These experimental
parameters can now be used to estimate the trends for free-hanging graphene under uniaxial strain.
Inserting $\gamma_{E_{2g}}$=1.99, $\beta_{E_{2g}}$=0.99, $\nu$=0.13 in Eq. 4, we get $\partial
\omega_{G^+}/\partial \varepsilon \sim$-18.6 cm$^{-1}$/ \%; $\partial \omega_{G^-}/\partial
\varepsilon \sim$-36.4 cm$^{-1}$/\%. Note that the effect of the substrate higher Poisson's ratio
is to significantly decrease the slope of the G$^+$ peak. These results are also in excellent
agreement with our first-principles calculations (see later).

We can now use our fitted $\gamma_{E_{2g}}$ to deduce the expected peak variations for graphene
under biaxial strain. In this case $\varepsilon_{ll}=\varepsilon_{tt}=\varepsilon $ and, from Eq.
3, $\Delta \omega_{E_{2g}} =-2 \omega^0_{E_{2g}}\gamma_{E_{2g}}\varepsilon$, since the shear
deformation term cancels. This means, as expected, that the G peak does not split. Also, no
difference is expected between free-hanging graphene and graphene on a substrate. Thus, for biaxial
strain: $\partial \omega_{G}/\partial \varepsilon \sim$-63 cm$^{-1}$/ \%.

To the best of our knowledge, no data exist in literature for uniaxial strain on graphite. However,
several authors applied hydrostatic pressure on graphite\cite{hanf,zhen,schi,windle} finding
$\partial \omega_{G}/\partial \sigma_h\sim$4.4-4.8cm$^{-1}$/GPa, where $\sigma_h$ is the
hydrostatic pressure (stress). The in-plane biaxial strain under hydrostatic pressure is
$\epsilon=(S_{ll}+S_{lt})\sigma_h$. Since for graphite in-plane $1/(S_{ll}+S_{lt})\sim
1/1250GPa~$\cite{bosak}, the data in Refs.\cite{hanf,zhen,schi,windle} correspond to an in-plane
Gruneisen parameter $\gamma_{E_{2g}}\sim$1.72-1.90, in very good agreement with our results. Many
groups have considered hydrostatic pressure on nanotubes (see,e.g.,\cite{Thomsen, Reich, windle}).
Generally it is found $\partial \omega_G/\partial \sigma_{h}\sim4-5 cm^{-1}/GPa$, in good agreement
with graphene and graphite. However, electron confinement and other effects in nanotubes warrant a
more detailed comparison of our results on graphene with the trends for the individual LO and TO G
bands in nanotubes, which will be presented elsewhere.

Several experiments exist for uniaxial strain on graphite fibres\cite{Galiotis}. These could be the
best approximation of uniaxial strain along the graphite plane, since their very large diameter
compared to single wall nanotubes ensures other possible effects due to electron confinement will
be negligible\cite{piscaPRB}. Extensive work on carbon fibres of different moduli has shown that
the peaks' shift is directly related to axial stress rather than strain\cite{melanitis}. Thus, one
can assume that in uniaxial experiments the applied stress is the known parameter, and the strain
applied to the atomic bonds can be derived from $\epsilon_{ll}=S_{ll}\sigma_{ll}$, where
S$_{ll}$=1/E is the fibre elastic compliance, E the Young's modulus and $\sigma_{ll}$ the applied
longitudinal stress. Thus, in order to correctly estimate the strain, it is necessary to know the
fibre E, which, in general, is significantly lower than the in-plane Young's modulus of
graphite\cite{Galiotis}. Then, if we extend the universal relation between Raman peak shift and
uniaxial stress to graphene, the following should hold: $\frac{\partial \omega_{Fibre}}{\partial
\varepsilon}$=$\frac {E_{Fibre}}{E_{Graphene}}\frac{\partial \omega_{Graphene}}{\partial
\varepsilon}$. Most fibres show a uniaxial stress sensitivity of $\partial \omega_G/\partial
\sigma_{ll}\sim2-3 cm^{-1}$/GPa~\cite{Galiotis}. In particular, PAN-based carbon fibres with "onion
skin" morphology (i.e. those most similar to large multi-wall nanotubes) have $\partial
\omega_G/\partial \sigma_{ll}$=-2.3 cm$^{-1}$/GPa~\cite{Galiotis}. Note that, due to disorder, the
G peak of carbon fibres is very broad and not resolved in two subbands. Thus, the fitted G
represents the average shift of the two subbands. Our average shift, using the in-plane graphite
Poisson's ratio, as needed in order to compare with fibres, is $\partial \omega_{G}/\partial
\epsilon \sim$-27cm$^{-1}/\%$. If we scale the uniaxial strain sensitivity of PAN fibres by the in
plane Young's modulus of graphite$\sim$1090GPa~\cite{bosak}, this would imply a value of
$\sim$-25cm$^{-1}/\%$, in excellent agreement with our average value. This also validates the
assumption that the graphene Young's modulus is similar to the in plane Young's modulus of
graphite, in agreement with recent measurements\cite{hone}. A notable discrepancy exists only with
Ref.\cite{Sakata} for uniaxial measurements on fibres. However, their data imply
$\gamma_{E_{2g}}\sim$2.87, in disagreement with both our measurements and with all graphite
literature\cite{hanf,zhen,schi,windle,Galiotis}. We also note that our results disagree with recent
Raman experiments on uniaxial strain in graphene\cite{ni,yu}, which report much smaller $\partial
\omega/\partial \varepsilon$, implying much smaller Gruneisen parameters. It is difficult to see
how the Gruneisen parameters of graphene should be much smaller than those measured in-plane for
graphite. Moreover, no G peak splitting was observed for uniaxial strain\cite{ni,yu}, again in
contrast with both our observation and general expectations.

We now consider the case of the singly degenerate phonon modes corresponding to the D and D' peaks.
The D peak is a breathing mode similar to the TO A$_{1g}$ phonon at K\cite{mapelli} (see Methods).
For a pure A$_{1g}$ symmetry and small strains, the uniaxial stress shift $\Delta\omega_{A_{1g}}$
is given only by the hydrostatic component of the stress:
\begin{equation}
\Delta\omega_{A_{1g}}=-\omega^0_{A_{1g}}\gamma_{A_{1g}}\left(\varepsilon_{tt}+
\varepsilon_{ll}\right)
\end{equation}
On the other hand, the D' phonon is of E symmetry\cite{mapelli} and we could expect in principle
splitting, and a relation similar to Eq. 4. However, experimentally this peak is very weak and we
cannot resolve any splitting in the strain range we have considered. Thus, for small strains, we
write for both Raman peaks
\begin{equation}
\Delta\omega_{2D;2D'}=-\omega^0_{2D;2D'}\gamma_{D;D'}\left(1-\nu\right)\varepsilon
\end{equation}

Combining our experimental data with Eq. 8 we get $\gamma_{D}\sim$3.55; $\gamma_{D'}\sim$1.61. For
free-hanging graphene, these give~$\partial
\omega_{2D}/\partial\varepsilon\sim$-83cm$^{-1}$/\%;~$\partial
\omega_{2D'}/\partial\varepsilon\sim$-45cm$^{-1}$/\%

In the case of graphene under biaxial strain $\varepsilon_{ll}=\varepsilon_{tt}=\varepsilon $ and
$\Delta \omega_{2D,2D'} =-2 \omega^0_{2D;2D'}\gamma_{D;D'} \varepsilon$. Thus, using our fitted
Gruneisen parameters, the expected 2D and 2D' variation as a function of biaxial strain are:
$\partial \omega_{2D} /\partial \varepsilon\sim$-191 cm$^{-1}$/ \% and $\partial \omega_{2D'}/
\partial \varepsilon \sim$-104 cm$^{-1}$/ \%.

To the best of our knowledge no data exist for the 2D or 2D' peak dependence in graphite as a
function of uniaxial strain. However, Ref.\cite{Galiotis} measured $\partial \omega_{2D}/\partial
\sigma_{ll}\sim 6.4cm^{-1}$/GPa for PAN carbon fibres. This scales to $\partial
\omega_{2D}/\partial \varepsilon\sim-70 cm^{-1}/\%$ in graphene, in agreement with our predicted
uniaxial trend, when using the in plane Possion's ratio of graphite, as needed for comparison with
fibres. For graphite under hydrostatic pressure Ref.\cite{goncha} reported $\partial
\omega_{2D}/\partial \sigma_h\sim$12.3 cm$^{-1}$/GPa, and $\partial \omega_{2D'}/\partial
\sigma_h\sim$9 cm$^{-1}$/GPa. This corresponds to an in-plane biaxial strain
$\epsilon=(S_{ll}+S_{lt})\sigma_h$. From $1/(S_{ll}+S_{lt})\sim 1/1250GPa$~\cite{bosak}, we get
$\partial \omega_{2D}/\partial \varepsilon\sim-154 cm^{-1}$/\%; $\gamma_{2D}$= 2.84; $\partial
\omega_{2D'}/\partial \varepsilon\sim-113 cm^{-1}$/\%; $\gamma_{2D'}$= 1.74, in broad agreement
with our predictions for biaxial strain.

Finally, we note that, in any case, the 2D peak is extremely sensitive to strain. With a typical
spectrometer resolution of $\sim$2cm$^{-1}$, a remarkable sensitivity of $\sim$0.01\% and 0.03\%
can be achieved for biaxial and uniaxial strain, respectively. We also note that a combined
analysis of G and 2D FWHM and shifts should allow to distinguish between effects of strain, doping
or disorder\cite{Pisana,ACFRamanSSC,DasCM,CasiraghiAPL,Yan}.
\begin{figure}
\centerline{\includegraphics [width=85mm]{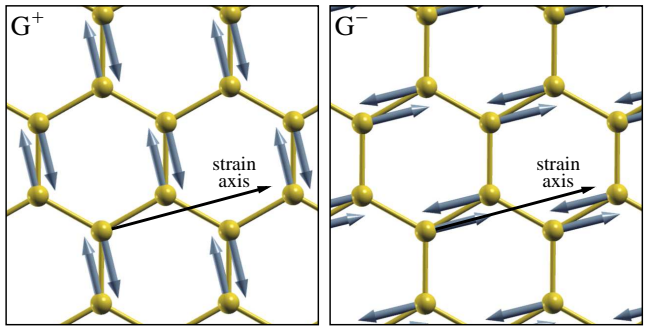}}
\caption{Eigenvectors of G$^+$ and G$^-$ modes as determined by density-functional perturbation theory. Note that these are perpendicular to each other, with G$^-$ polarized
along the strain axis, as expected}
\label{Figure4}
\end{figure}

To further understand our findings we perform first-principles calculations on free-standing
graphene as described in Methods\cite{mounet}, for small strains up to $\sim$1\%, to compare with
experiments. The effects on electron and phonon bands of larger strains will be reported elsewhere.
Fig.4 plots the resulting G$^+$/G$^-$ eigenvectors. These are perpendicular to each other, with the
G$^-$ eigenvector oriented along the strain direction, as expected. For small strains we find
$\partial \omega_{G^-}/\partial \varepsilon \sim$-34 cm$^{-1}$/ \% and $\partial \omega_{G^+}/
\partial \varepsilon \sim$-17 cm$^{-1}$/\%, independent on the strain direction, as expected from
symmetry. We also get $\gamma_{E_{2g}}$=1.87; $\beta_{E_{2g}}$=0.92, in excellent agreement with
our measured parameters. Note that, in order to compare the calculated trends for G$^+$ and G$^-$
with our measurements, we need to insert the theoretical parameters in Eq. 4 together with the
substrate Poisson's ratio. This gives $\partial \omega_{G^-}/\partial \varepsilon
\sim$-30cm$^{-1}$/ \%; $\partial \omega_{G^+}/\partial \varepsilon \sim$-10.3cm$^{-1}$/\%, in
excellent agreement with the fits in Fig.3a. We also calculate the biaxial strain variation for the
G peak. We find $\partial \omega_{G}/\partial \varepsilon \sim$-58
cm$^{-1}$/\%;~$\gamma_{E_{2g}}$=1.8, again in excellent agreement with the biaxial values based on
our experimental Gruneisen.

We then calculate the uniaxial and biaxial strain variation for the 2D peak. We find $\partial
\omega_{2D} /\partial \varepsilon\sim$-60 cm$^{-1}$\% for uniaxial, and $\partial \omega_{2D}
/\partial \varepsilon\sim$-144 cm$^{-1}$/ \%; for biaxial and $\gamma_D\sim$2.7 for both. These are
in excellent agreement with the results of hydrostatic pressure experiments on graphite, and in
broad agreement with our experimental data for uniaxial strain (and the consequent biaxial
predictions), being $\sim$25/\% smaller. It is important to consider that, while for the Raman
active G mode we are probing the same centre-zone phonon when measuring the Raman spectrum on a
strained sample, the Raman D and D' peaks are zone boundary phonons activated by double resonance
(see Methods). Any change in the double Resonance condition during the strain experiments will vary
the actual phonon probed in the Raman measurements, as well as inducing a change in the phonon
frequencies. Thus, the relationship between phonon Gruneisen parameters and the variation of the
Raman peaks with applied strain is in principle more complex than the case of the G peak and what
implied by Eqs~7,8. Indeed, while biaxial strain does not move the relative positions of the Dirac
cones, uniaxial strain changes them~\cite{yang}. Note that this does not open any gap, in contrast
with the conclusions of Ref.\cite{ni}. Still, it can have a significant influence in the double
resonance process. As explained in Methods, while the D' is intra-valley, i.e. connecting two
points belonging to the same cone around \textbf{K} or \textbf{K'}, the D peak phonon requires
scattering from the cone around \textbf{K} to that around
\textbf{K'}\cite{steffi2,PiscanecPRL,ACFRaman}. Thus, its wavevector is determined by the relative
distance of the Dirac cones and by the laser excitation energy. Our experiments are performed for a
fixed excitation. Then, what we measure in Raman spectroscopy of uniaxially strained graphene is
the combination of the 2D phonon shift due to strain, and a possible additional shift due to the
fact that the relative movement of the Dirac cones changes the phonon wavevector. For an asymmetric
movement this could lead to peak broadening and splitting. Indeed the experimental FWHM(2D)
significantly increases with strain. In the case of the 2D' peak the movement of the relative
positions of the cones will have no consequence, since it is an intra-valley process. However, for
both D and D', other effects could be given by the renormalisation of Fermi velocity and phonon
group velocity with strain. Thus, especially for the D peak, our measured $\gamma_{D}$ has to be
taken as an upper boundary, and a more general expression to evaluate it can be $\gamma_{D} =
-\frac{\Delta \omega_{2D}-\Delta' \omega_{2D}} {\omega^0_{2D}\left(1-\nu\right)\varepsilon}$, with
$\Delta' \omega_{2D}$ encompassing the corrections due to the changes in the phonon selected in
double resonance, as a function of strain. We note that, in the case of biaxial strain, at least
the effects due to the relative movement of the Dirac cones are absent. Thus, Raman experiments on
graphene under biaxial strain would be more suited to measure the D mode Gruneisen parameter, and
this explains why our calculations are in excellent agreement with the hydrostatic pressure
experiments on graphite. Thus, given the peculiar nature of electron-phonon and electron-electron
interactions around the K point in graphene\cite{PiscanecPRL,Pisana,BaskoBig}, combined with the
relative movement of the K, K' points under uniaxial strain\cite{yang}, and the possible
re-normalizations of electron and phonon bands, the full theoretical description of the 2D peak
under uniaxial strain needs further investigation, and will be reported elsewhere.
\begin{figure}
\centerline{\includegraphics[width=90mm]{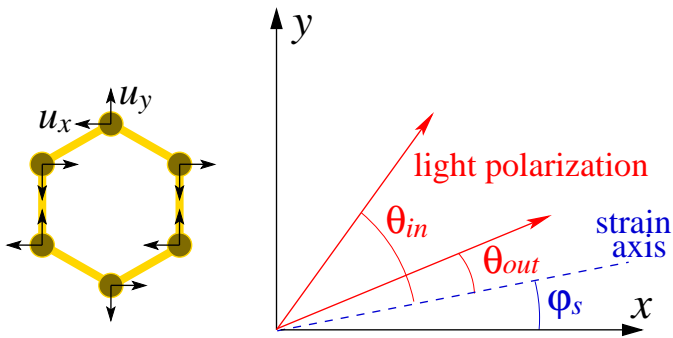}}
\caption{\label{fig:axes} (Color online.) Geometry implied by Eqs. (\ref{Hint=})-(\ref{IlIt=}).
The circles in the hexagon represent carbon atoms.
The $x$~axis is chosen to be perpendicular to the C-C bond.
The short black arrows represent the phonon displacements in the
$(x,y)$ basis, as assumed in Eq.(\ref{Hint=}) (the longitudinal and
transverse normal modes are given by their linear combinations).
The strain axis is the blue dashed line. The red arrows represent the polarization of incident and detected light.}
\label{figure5}
\end{figure}
\begin{figure*}
\centerline{\includegraphics[width=150mm]{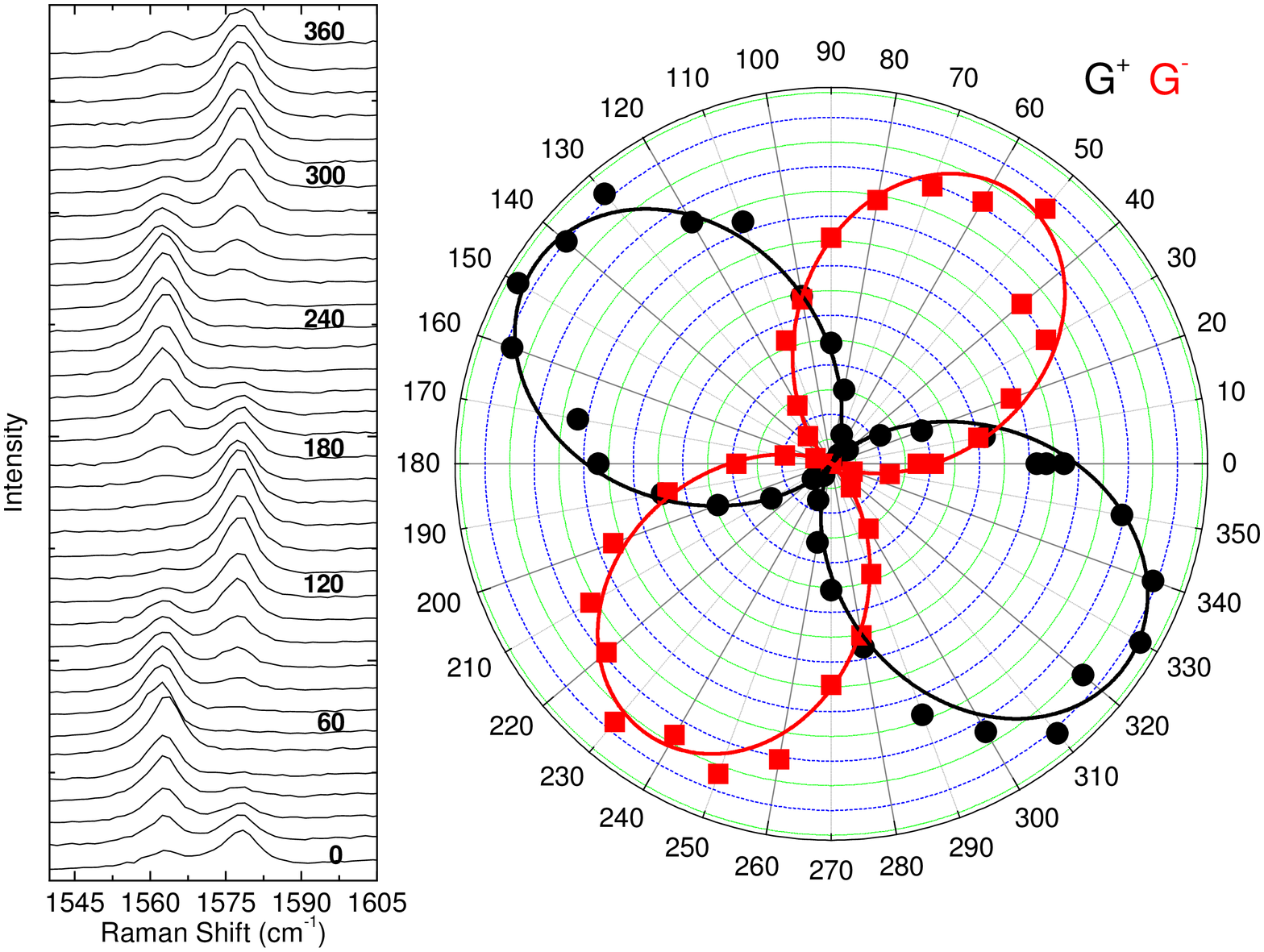}} \caption{(Color online.) (Left) Raman spectra
and (right) polar plot of the fitted G$^+$ and G$^-$ peaks as a function of the angle between the
incident light polarization and the strain axis $\theta_{in}$, measured with an analyzer selecting
scattered polarization along the strain axis, $\theta_{out}=0$. The polar data are fitted to
$I_{G^-}\propto\cos^2(\theta_{in}-56^\circ)$ and $I_{G^+}\propto\sin^2(\theta_{in}-56^\circ)$.}
\label{figure6}
\end{figure*}

We now consider the polarization dependence of the G$^+$ and G$^-$ intensities, expected due to the
nature of the phonon eigenvectors and their orientation with respect to the strain~\cite{Sakata}.
The effective photon-phonon interaction Hamiltonian for the $E_{2g}$ phonons is~\cite{BaskoBig}:
\begin{equation}\label{Hint=}
H_{int}\propto\left[(\mathcal{E}_x^{in}\mathcal{E}_x^{out}
-\mathcal{E}_y^{in}\mathcal{E}_y^{out})u_x
+(\mathcal{E}_x^{in}\mathcal{E}_y^{out}+
\mathcal{E}_y^{in}\mathcal{E}_x^{out})u_y\right]
\end{equation}
Here $\mathcal{E}_x^{in(out)},\mathcal{E}_y^{in(out)}$ are the cartesian components of the electric
field of the incident (scattered) light, and $u_x,u_y$ are the phonon displacements in the
$(x,y)$-basis (see Fig.~\ref{figure5} for details). The $x$ axis is chosen perpendicular to the
C--C bond. This Hamiltonian is the only allowed by the $C_{6v}$ symmetry of graphene. In the
presence of strain the Hamiltonian changes, but the correction will be of the order of the strain
itself. For a fixed small strain, these corrections can be ignored, in first approximation, in the
calculation of the polarization dependence of the G bands. The main effect of strain is to force
the phonon normal modes to be longitudinal ($u_l$) and transverse ($u_t$) with respect to the
strain axis, as discussed above, and shown in Fig. 4. If we call $\varphi_s$ the angle between the
strain axis and the $x$~axis, we can write:
\begin{equation}
u_x=u_l\cos\varphi_s+u_t\sin\varphi_s,\quad
u_y=-u_l\sin\varphi_s+u_t\cos\varphi_s
\end{equation}
In our Raman spectrometer, we can excite with linearly polarized light and use an analyzer for the
scattered radiation. This means that the corresponding electric field vectors have definite
orientations: $\mathcal{E}_x^{in,out}=\mathcal{E}_0^{in,out}\cos(\theta_{in,out}+\varphi_s)$,
$\mathcal{E}_y^{in,out}=\mathcal{E}_0^{in,out}\sin(\theta_{in,out}+\varphi_s)$, where the
polarization is measured with respect to the strain axis. Substituting these in Eq.~(\ref{Hint=}),
the matrix elements corresponding to emission of longitudinal and transverse phonons are
proportional to $\cos(\theta_{in}+\theta_{out}+3\varphi_s)$ and
$\sin(\theta_{in}+\theta_{out}+3\varphi_s)$, respectively. The intensities of the two peaks are
given by their squares:
\begin{equation}\label{IlIt=}
I_{G^-}\propto\cos^2(\theta_{in}+\theta_{out}+3\varphi_s),\quad
I_{G^+}\propto\sin^2(\theta_{in}+\theta_{out}+3\varphi_s)
\end{equation}

To test this we do polarization measurements with an analyzer for the scattered light aligned with
the strain direction($\theta_{out}=0$), and rotating the incident polarization with respect to the
strain axis in steps of 10$^\circ$, Fig.6. The data in Fig. 6 are well fitted by
$I_{G^-}\propto\cos^2(\theta_{in}-56^\circ)$ and $I_{G^+}\propto\sin^2(\theta_{in}-56^\circ)$.
According to Eq.~(\ref{IlIt=}), this gives $\varphi_s=-18.7^\circ$. We thus get the orientation of
the graphene crystal with respect to the known strain axis.

The physical origin of the polarization dependence of the $G^+/G^-$ peaks can be traced to the
microscopic mechanism of Raman scattering. The light interaction with graphene phonons is mediated
by electrons. As discussed in\cite{BaskoBig} for unstrained graphene, if one assumes the electron
spectrum to be isotropic (Dirac), the G peak intensity vanishes. Thus, the G peak is entirely due
to the anisotropic terms in the electronic spectrum. In other words, in order to contribute to the
G peak, electrons must ``feel'' the crystallographic directions. In unstrained graphene this has no
consequence, since the two vibrations are degenerate and not resolved. Under strain, the two
sub-bands correspond to definite orientations of the vibrations with respect to the strain axis. It
is thus the interaction of electrons, which ``feel'' the crystallographic directions, with phonons,
entirely determined by the the strain direction, that gives the polarization dependence.

In summary, we probed with Raman spectroscopy the optical phonons of graphene as a function of
uniaxial strain. We find that the doubly degenerate E$_{2g}$ mode splits in two components, one
polarized along the strain, the other, perpendicular. This split of the Raman G peak in 2 subbands,
G$^+$ and G$^-$, is analogous to that induced by curvature in nanotubes. These subbands red-shift
with increasing strain, whilst their splitting increases, in excellent agreement with
first-principles calculations. Their relative intensities vary with polarization, allowing to probe
the sample crystallographic orientation with respect to the strain. The 2D and 2D' bands downshift,
but do not split for small strains. Our results can be used to quantify the amount of uniaxial or
biaxial strain, providing a fundamental tool for graphene-based nanoelectronics and nano/micro
electro mechanical systems.

\textbf{Acknowledgements} ACF, KSN, AKG acknowledge funding from the Royal Society and the European
Research Council; AL from the University of Palermo, Sicily; TM form the Sultan Qaboos University;
NB and NM from the Interconnect Focus Center, a Semiconductor Research Corporation program, and
from MITRE.

\section{Methods}
\subsection{Strain and Raman Measurements}
In order to controllably and reproducibly induce strain, graphene layers, prepared by
micromechanical cleavage of graphite, are deposited on two different flexible substrates. One is a
720 $\mu m$ thick, 23mm long Polyethylene terephthalate (PET) film.  The other is a 3 mm thick, 10
cm long, 1 cm wide clear acrylic (Perspex). In both cases the large length-to-width ratio is chosen
to allow uniform bending and reversibility. Prior to graphene deposition, the substrates are spin
coated with SU8 2000.5 (MicroChem) photoresist\cite{su8} of carefully chosen thickness (400nm),
which is then cross-linked. This ensures optimal visible contrast for graphene
identification\cite{blakeAPL,CasiraghiNL}. To achieve maximum strain, the length of the substrate
is altered in order to have the flake at its center, Fig. 1. Note that the size of the graphene
layers is orders of magnitude smaller than the substrate length ($\sim10^3$ and $\sim10^4$ times
smaller, respectively). This ensures a uniform strain in the section measured by Raman
spectroscopy. The first substrate is used in two point bending experiments, whilst the second in
four point bending, Fig. 1. Raman spectra are measured with a 100X objective at 514nm excitation
with a Renishaw micro-Raman spectrometer, having 1800 grooves/mm grating and spectral resolution of
$\sim$2cm$^{-1}$. The polarization of the incident light can be controlled by a Fresnel rhomb,
while an analyzer can be placed before the grating. The power on the samples is well below 2mW, so
that no shift, nor change in width of the Raman peaks is observed for a fixed strain, thus ensuring
no damage, nor heating. A cycle of loading, unloading and loading is followed to ensure
reproducibility for both experiments. A total of 80 Raman spectra are measured for an average
strain increment of 0.05\%. The maximum strain applied to the sample is less than $\sim$1.2\%. In
the two point measurements, the spectra do not change until a nominal strain of $\sim$0.55\% is
applied to the substrate. Afterwards they evolve linearly with strain. Thus, we assume this point
as the reference zero strain for the sample. In the four point, the spectra evolve linearly from
zero strain. The two set of data are fully overlapping, further confirming the strain measurements.
The data are fully reproducible over three strain cycles between maximum and minimum, as shown in
Fig. 3. Only when suddenly applying large strains or large strain increments we observe sample
slippage, indicated by an upshift, or smaller downshift or no shift at all of the Raman parameters.
Indeed, for samples suddenly bent to large strain values of a few \% we often observe no change in
the Raman peaks, indicating a general loss of contact between the graphene and the substrate.
\textit{It is thus extremely important to apply the strain in the most controlled way in order to
ensure reproducibility and no slippage}. A further set of 36 measurements is done for a fixed value
of strain, by rotating the incident polarization in 10$^\circ$ steps with respect to the strain
axis, and analyzing the scattered light in the plane parallel to the strain axis.
\subsection{Origin of the Raman peaks}
All carbons show common features in their Raman spectra in the 800-2000 cm$^{-1}$ region, the
so-called G and D peaks, which lie at around 1580 and 1350 cm$^{-1}$ respectively\cite{acfRS}. The
G peak corresponds to the doubly degenerate $E_{2g}$ phonon at the Brillouin zone center. The D
peak is due to the breathing modes of sp$^2$ rings and requires a defect for its
activation\cite{tuinstra,ramanprb}. It comes from TO phonons around the \textbf{K} point of the
Brillouin zone\cite{tuinstra,ramanprb}, is active by double resonance (DR)\cite{bara,steffi2} and
is strongly dispersive with excitation energy due to a Kohn Anomaly at
\textbf{K}\cite{PiscanecPRL}. The activation process for the D peak is an inter-valley process as
follows: i) a laser induced excitation of an electron/hole pair; ii) electron-phonon scattering
with an exchanged momentum $\textbf{q}\sim\textbf{K}$; iii) defect scattering; iv) electron/hole
recombination. The D peak intensity is not related to the number of graphene layers, but only to
the amount of disorder\cite{tuinstra,ramanprb}. DR can also happen as intra-valley process i.e.
connecting two points belonging to the same cone around $\textbf{K}$ (or $\textbf{K}'$). This gives
rise to the so-called D'peak, which can be seen around 1620 cm$^{-1}$ in defected graphite
\cite{nemanich}. The 2D peak is the second order of the D peak. This is a single peak in monolayer
graphene, whereas it splits in four bands in bilayer graphene, reflecting the evolution of the band
structure\cite{ACFRaman}. The 2D' peak is the second order of the D' peak. Since 2D and 2D' peaks
originate from a process where momentum conservation is obtained by the participation of two
phonons with opposite wavevectors ($\textbf{q}$ and $-\textbf{q}$), they do not require the
presence of defects for their activation, and are thus always present. Indeed, high quality
graphene shows the G, 2D and 2D' peaks, but not D and D'\cite{ACFRaman}.
\subsection{Density Functional Calculations}
We use density-functional theory (DFT) and density-functional perturbation theory
(DFPT)~\cite{dfpt} as implemented in the {\sc PWSCF} package of the {\sc Quantum-ESPRESSO}
distribution~\cite{pwscf}, within the local-density approximation~\cite{lda}, with norm-conserving
pseudopotentials~\cite{pseudi} and a plane-wave expansion up to 55~Ry cut-off. The Brillouin-zone
is sampled on a 42$\times$42$\times$1 Monkhorst-Pack mesh for graphite and graphene, with a cold
smearing\cite{marzari99} in the electronic occupations of 0.02~Ry.  We use the equilibrium lattice
parameter $a$ = 2.43~\AA~and an interlayer spacing of 15~\AA. We apply the strain in different
directions. For each direction and strain we determine the structure with the lowest total energy,
by varying the size of the unit cell in the direction perpendicular to the strain. Our calculated
values at zero strain are $\omega_{G_0}$=1603.7 cm$^{-1}$, $\omega_{D_0}$=1326 cm$^{-1}$ and
$\nu$=0.15.

\end{document}